\documentclass[twocolumn,showpacs,preprintnumbers,nofootinbib,prd,superscriptaddress,groupedaddress,10pt]{revtex4-1}

\usepackage[centertags]{amsmath}
\usepackage{amssymb}
\usepackage{latexsym}
\usepackage{enumerate}
\usepackage{graphicx}
\usepackage{mathrsfs}
\usepackage{hyperref}
\usepackage{stmaryrd}
\usepackage{import}
\usepackage{tensor}
\usepackage{color}
\usepackage{bm}
\usepackage{multirow}
\usepackage{float}

\newcommand{\bi}{\begin{itemize}}
\newcommand{\ei}{\end{itemize}}
\newcommand{\be}{\begin{equation}}
\newcommand{\ee}{\end{equation}}

\renewcommand{\l}{\left(}
\renewcommand{\r}{\right)}

\renewcommand{\d}{\delta}

\newcommand{\La}{\Lambda}
\newcommand{\la}{\lambda}

\renewcommand{\o}{\omega}
\renewcommand{\th}{\theta}

\newcommand{\q}{\quad}

\newcommand{\vp}{\varphi}

\newcommand{\pa}{\partial}

\begin{document}

\title{Unbound motion on a Schwarzschild background: \\
Practical approaches to frequency domain computations}
\author{Seth Hopper}
\affiliation{CENTRA, Departamento de F\'{\i}sica, Instituto Superior T\'ecnico -- IST, Universidade de Lisboa -- UL,
Avenida Rovisco Pais 1, 1049 Lisboa, Portugal}

\begin{abstract}
Gravitational perturbations due to a point particle moving on
a static black hole background are naturally
described in Regge-Wheeler gauge. The first-order field equations
reduce to a single master wave equation for each radiative mode.
The master function satisfying this wave equation is a linear combination 
of the metric perturbation amplitudes with a source term arising
from the stress-energy tensor of the point particle. The original
master functions were found by Regge and Wheeler (odd parity)
and Zerilli (even parity). Subsequent work by Moncrief and then
Cunningham, Price and Moncrief introduced new master variables
which allow time domain reconstruction of the metric perturbation
amplitudes. Here I explore the relationship between these different
functions and develop a general procedure for deriving new
higher-order master functions from ones already known. The benefit of
higher-order functions is that their source terms always converge
faster at large distance than their lower-order counterparts. 
This makes for a dramatic improvement in both the speed and
accuracy of frequency domain codes when analyzing unbound
motion.
\end{abstract}

\maketitle

\section{Introduction}
\label{sec:intro}

The recent first detections \cite{AbboETC16a,AbboETC16b,AbboETC17} 
of gravitational waves 
were made possible, in part, by accurate modeling of strongly
gravitating binary sources. Through a mixture of numerical relativity,
post-Newtonian theory, and black hole perturbation theory methods,
the inspiral-merger-ringdown waveform is being modeled with increasing accuracy
e.g.~\cite{BuonDamo99,BoheETC2016,KhanETC15}. 
This work on bound motion has been incredibly important, 
but in the decades while these techniques have been
developed, the modeling of unbound binary sources has been
largely neglected. 

In this work and its companion paper,
Ref.~\cite{HoppCard17} with Cardoso, we return focus to unbound motion. 
In Ref.~\cite{HoppCard17} (and also earlier in Ref.~\cite{CardETC16}) we provide numerical results from scattering systems,
which we compare to various analytical predictions. Here, I describe
the numerical method used to generate those results.

In particular,
I consider point particle motion in Schwarzschild
spacetime, be it a plunge or a scattering event. Finding the metric
perturbation (MP) sourced by such a particle requires solving the first-order
field equations. As usual, this task requires specifying a first-order gauge 
condition and subsequently making several choices concerning numerical techniques.
Perhaps the most natural approach would be to decompose the system in spherical 
harmonic modes and then work in Lorenz gauge 
with a 1+1 time domain (TD) solver. Such an approach has been employed by 
Barack  and Sago \cite{BaraSago10} for eccentric motion. TD codes have 
the advantage 
of being `source agnostic', in that the method need not change depending on
the type of motion (bound, unbound, circular). On the other hand, TD codes
have the problem of junk radiation due to unphysical initial data 
\cite{FielHestLau10}. And, while for bound motion junk radiation will die off
as the system `relaxes' into the desired solution, a particle on
an unbound trajectory will chase the wavefront of the junk radiation,
forcing a TD simulation to begin far from the black hole, 
especially for high energy events.
Additionally, for eccentric motion, TD codes have not been able to compete with
the speed and accuracy of frequency domain (FD) approaches 
\cite{AkcaWarbBara13,OsbuETC14,vandShah15} (even for eccentricities approaching 0.8).

This advantage of FD techniques stems largely from the 
ease of solving a set ordinary
differential equations (ODEs) 
rather than evolving a partial differential equation.
Further, while generic source motion would have to be decomposed in the 
FD with a Fourier transform, 
bound geodesics brings the benefit of exactly periodic 
radial motion. Thus, the spectral decomposition of the source and its 
field are represented by Fourier series with exactly known frequencies.
From this follows the exponential convergence of source integration 
\cite{HoppETC15}, further improving the speed and accuracy of FD codes.

The success of FD methods for bound motion leads one to ask whether the techniques
can be usefully generalized to unbound motion. The first challenge faced is
the loss of the discrete spectrum provided by a Fourier series. This in itself
is not an immediate reason to despair; certainly,
spectral techniques are used across all areas of science
to analyze non-periodic systems, usually
in the form of discrete Fourier transforms.
One must simply choose the largest and smallest frequencies to consider in the 
problem. Fundamentally, this is no different than choosing the smallest time step
in a TD evolution and the total length of evolution.

The second challenge comes from considering a source which has traveled 
all the way from spatial infinity. Experience and physical intuition tell us
that the `interesting part' of the particle's motion happens when it is
(in some sense) near the black hole. 
Even a particle with a large Lorentz factor should simply behave as if it
were in Minkowski spacetime when it is far away. 
Mathematically, this intuition is born out by examining the large-$r$
behavior of sources.
In order to transform
the TD sources to their FD amplitudes, one must (formally) 
integrate over all time. For such an integral to converge, the source
must fall off at least a $1/r$, and indeed, 
examining the Lorenz gauge sources, one finds that this is always true.

And yet, of course Lorenz is not the only gauge choice available.
It is often preferable to work in Regge-Wheeler (RW) gauge on Schwarzschild
spacetime. Then, by employing the Regge-Wheele-Zerilli (RWZ)
formalism the full field equations can be reduced to a single
`master' wave equation for each $\ell,m$ mode. While there appears to be little
downside (at least when computing fluxes) 
to this approach when considering bound motion, the source terms to
the master equation are not as well behaved as the Lorenz sources.
The main result of this paper is to show how the RWZ sources can be 
modified so as to improve their large-$r$ behavior, thus
improving the speed and accuracy of codes written to analyze unbound
motion around a static black hole.

The numerical study of unbound point particle motion on Schwarzschild
spacetime has a long history extending back to pioneering work by
Davis et al.~\cite{DaviETC71}. They used the Zerilli
equation \cite{Zeri70} to compute the radiated energy due to a particle falling
head-on into a Schwarzschild black hole from rest at infinity.
Of course, both before and since Ref.~\cite{DaviETC71}, a variety of analytic
techniques, e.g.~\cite{Pete70,Turn77,KovaThor78,BlanScha89}, have been
applied to the problem. We compare to many of those predictions 
in Ref.~\cite{HoppCard17}, but they will not be considered here.

A great breakthrough in black hole perturbation theory was made by Teukolsky
who derived his eponymous equation \cite{Teuk73} using the Newman-Penrose formalism.
The Teukolsky equation describes scalar, neutrino, electromagnetic, and gravitational
perturbations on a Kerr black hole background all in one master equation.
Detweiler and Szedenits  \cite{DetwSzed79} made use of Teukolsky's equation 
(along with a shrewd intuition for which divergent integrals to ignore) to 
analyze plunges with nonzero angular momentum on a Schwarzschild background.

While the Teukolsky formalism is indeed powerful, the consolidation of so many
different phenomena into one equation has its costs. Two of these costs,
a radial equation with a long-ranged potential and poorly behaving source terms
were rectified by the transformations introduced by Sasaki and Nakamura \cite{SasaNaka82}.
(Subsequently, the poor source term behavior was explained by Poisson \cite{Pois96} 
and Campanelli and Lousto \cite{CampLous97}.)
The Sasaki-Nakamura equation has been used extensively to study
unbound motion. In particular, work by 
Oohara \cite{Ooha83}, Oohara and Nakamura \cite{OohaNaka83},
and, more recently, Berti et al.~\cite{BertETC10} considered the special
case of static black holes.

While the Sasaki-Nakamura formalism is very powerful, I will largely ignore 
it here. It remains invaluable for studying point-particle motion on Kerr
spacetime (especially for unbound sources), but is needlessly complicated
in the Schwarzschild case. Obtaining source terms for the Sasaki-Nakamura 
equation requires solving an additional numerical integral, which we 
are able to sidestep entirely through the methods described here.

Before concluding this introduction, it is worth mentioning the 
relation of this work to gravitational self-force (GSF) research. 
Recent black hole perturbation theory methods and codes have been developed
in no small part because of the desire to crack the GSF problem.
The goal is to model the motion of a small-mass particle on a Kerr background
while using effect of the particle's own MP to drive it off the 
background geodesic. GSF research is important
for the eventual detection of extreme mass-ratio binaries by LISA \cite{LISA}, 
or a similar detector. The prospect of computing the GSF for an unbound
trajectory and finding, e.g., the correction to the particle's deflection
angle is tantalizing. But, it is beyond the scope of this paper, and I will 
not attempt to compute the GSF here, settling for the more modest goal 
of developing a reliable method for finding waveforms and fluxes, which
are interesting in their own right.

This paper is organized as follows. Sec.~\ref{sec:statement}
explains the mathematical basis for the problem this paper sets out to solve.
In Sec.~\ref{sec:masterDeriv} I explore the relationship between
known master functions and also show how to generalize this relationship to define 
master functions with well-behaved source terms. Sec.~\ref{sec:results}
gives an overview of the numerical algorithm I have implemented and discusses
the practical benefits of the techniques developed here.
In Sec.~\ref{sec:prospects} I explore whether the methods presented
in this paper would work in other systems. Then, the Appendix gives
details of unbound geodesics, the RWZ formalism, and specific source terms.
Throughout this paper I let $G = c = 1$ and use standard Schwarzschild
coordinates $x^\mu = (t, r, \th, \vp)$. 
A subscript $p$ indicates a field evaluated at the particle's location,
e.g.~$r_p = r_p (t)$.

\section{Statement of the problem}
\label{sec:statement}

\subsection{Quick background of the RWZ formalism}
\label{sec:rwzBackground}

I now consider the process of solving the 
first-order (in mass ratio) RW gauge field equations in the FD. 
As the general formalism has now been well established, I relegate its
full presentation to App.~\ref{sec:rwzFull}. For the purposes of this section it 
is enough to recall the following facts. 
The master equation in the TD is of the form 
\begin{align}
\begin{split}
\label{eq:TDmastereqn}
& \left[-\frac{\pa^2}{\pa t^2} + \frac{\pa^2}{\pa r_*^2} - V_\ell(r)\right]
\Psi_{\ell m}(t,r) = S_{\ell m}(t,r), 
\end{split}
\end{align}
where $r_* = r + 2M \log (r/2M - 1)$ is the usual tortoise coordinate.  
Both the potential $V_\ell$ and the source term $S_{\ell m}$  
are ($\ell + m$) parity dependent.
The master function $\Psi_{\ell m}$ and its source $S_{\ell m}$ are 
decomposed into harmonics $X_{\ell m \omega}$ 
and $Z_{\ell m \omega}$  using a Fourier transform. 
They satisfy the FD version of 
Eq.~\eqref{eq:TDmastereqn},
\be
\label{eq:FDmastereqn}
\left[\frac{d^2}{dr_*^2} +\o^2 -V_\ell(r)\right]
X_{\ell m\o}(r) = Z_{\ell m\o}(r).
\ee
It is typical to solve this equation by the method 
of variation of parameters, i.e.~finding its homogeneous solutions
(denoted $\hat X^\pm_{\ell m \omega}$, with the $+$ solution
being outgoing as $r_* \to \infty$ and $-$ being downgoing 
as $r_* \to -\infty$)
and then integrating them against the source,
which has the specific distributional form 
$S_{\ell m}(t) \equiv G_{\ell m}(t) \, \delta (r - r_p) + F_{\ell m}(t) \,\delta' (r - r_p).$
From a practical standpoint, the crux
of the numerical calculation amounts to solving the 
integral \cite{HoppEvan10},
\begin{align}
\label{eq:EHSC}	
&C_{\ell m \omega}^\pm  =  \frac{1}{W_{\ell m\o}} \int_{-\infty}^{\infty}
\Bigg[ 
 \frac{1}{f_{p}} \hat X^\mp_{\ell m \omega} (r_p)
 G_{\ell m} (t) \hspace{5ex} \\ \notag
&\hspace{1ex} 
+ \l \frac{2M}{r_{p}^2 f_{p}^{2}} \hat X^\mp_{\ell m \omega} (r_p)
 - \frac{1}{f_{p}} 
 \frac{d \hat X^\mp_{\ell m \omega}(r_p)}{dr} \r F_{\ell m} (t)
 \Bigg]  e^{i \o t}  \, dt ,
\end{align}
where $f \equiv 1 - 2M /r$ and $W_{\ell m \omega}$ is the Wronskian.
The constants $C^\pm_{\ell m \omega}$ are the \emph{normalization coefficients},
which must be found
for a range of spherical harmonic indices $\ell,m$ and frequencies $\omega$.

For eccentric geodesic motion, the normalization coefficients, 
along with homogeneous solutions to Eq.~\eqref{eq:FDmastereqn}, define the 
\emph{extended homogeneous solution} (EHS) \cite{BaraOriSago08}, which is the
correct TD solution to Eq.~\eqref{eq:TDmastereqn}. 
The EHS method provides exponential convergence of Fourier harmonics
everywhere, including the particle's location, making it critical to 
fast and accurate FD GSF codes. However, in the present
context one of the crucial assumptions of EHS fails, namely the presence
of a source-free region where the Fourier synthesis is known to converge
exponentially. As such, the EHS method is not directly applicable to
unbound motion and it remains to be seen if the method can be 
suitably altered to once again provide exponential convergence and 
avoid the Gibbs behavior that results from using a singular source.
For this work I do not pursue the local reconstruction of the MP, but 
focus rather on efficient methods for solving Eq.~\eqref{eq:EHSC}.
These normalization coefficients
are all that are needed to compute the waveform and the  
total radiated energy and angular momentum.

\subsection{RWZ source behavior for various master functions}

For concreteness, imagine a particle plunging from spatial infinity.
It is clear that for the integral \eqref{eq:EHSC} to converge both $G_{\ell m}$ 
and $F_{\ell m}$ must fall off at least as 
$1/r_p$ far from the black hole (when $t \to - \infty$).
At the horizon (when $t \to \infty$), $G_{\ell m}$ must fall off at least as $f_p^2$ while
$F_{\ell m}$ must fall off at least as $f_p^3$.
(Note that $\hat X_{\ell m \omega}^\pm  \sim e^{\pm i \omega r_*}$ 
in the asymptotic regimes, so this does not help convergence.) 

Consider now the Zerilli (Z) \cite{Zeri70}  source,
used by Davis et al.~\cite{DaviETC71}. 
(They used the axial symmetry of their problem, and so
their source looks simpler than the generic 
source shown in App.~\ref{app:Sources}.)
Expanding at the horizon we see,
\begin{align}
G_{\ell m}^{Z} \sim f_p^2 + \mathcal{O} \left( f_p^3 \right), 
\q
F_{\ell m}^{Z} \sim f_p^3 + \mathcal{O} \left( f_p^4 \right), 
\end{align}
as $r_p \to 2M.$
These converge fast enough; in fact, all master function 
source terms fall off sufficiently fast
at the horizon, so I will not consider their expansions further.
At large $r_p$
\begin{align}
\label{eq:ZTrend}
G_{\ell m}^{Z} \sim r_p^{-2} + \mathcal{O} \left( r_p^{-3} \right), 
\q
F_{\ell m}^{Z} \sim r_p^{-1} + \mathcal{O} \left( r_p^{-2} \right).
\end{align}
As expected, this source satisfies the necessary requirements for the convergence
of the integral in Eq.~\eqref{eq:EHSC}. In the odd-parity sector, the original
variable is due to Regge and Wheeler \cite{ReggWhee57}. 
Expanding its source at large $r_p$,
\begin{align}
\begin{split}
G_{\ell m}^{\rm RW} &\sim r_p^{-3} + \mathcal{O} \left( r_p^{-4} \right), 
\q
F_{\ell m}^{\rm RW} \sim r_p^{-3} + \mathcal{O} \left( r_p^{-4} \right).
\end{split}
\end{align}
 These terms, too, trend 
so that the normalization integral converges.

Recently, the original RW and Z master functions are less used
because they do not permit TD reconstruction of the MP amplitudes. It is more common
to use the Zerilli-Moncrief (ZM) \cite{Monc74}  
and Cunningham-Price-Moncrief (CPM) \cite{CunnPricMonc78}  functions
which do allow for MP reconstruction in the TD (see \cite{MartPois05,HoppEvan10}). 
The more recent ZM and CPM variables are (\emph{almost} --- see next section) 
the time integrals
of the Z and RW variables.
For bound motion the ZM and CPM variables are preferable in every way
to the Z and RW variables. However, in the unbound case the 
ZM source term behaves poorly at large $r_p$,
\begin{align}
\begin{split}
\label{eq:ZMTrend}
G_{\ell m}^{\rm ZM} &\sim r_p^{-1} + \mathcal{O} \left( r_p^{-2} \right), 
\q
F_{\ell m}^{\rm ZM} \sim r_p^{0} + \mathcal{O} \left( r_p^{-1} \right).
\end{split}
\end{align}
The $F^{\rm ZM}_{\ell m}$ term prevents 
the normalization integral \eqref{eq:EHSC} from converging. Meanwhile the CPM
variable still converges at large $r_p$ implying that it can be 
used to analyze unbound sources, but is less
effective than the RW variable since it falls off significantly more slowly,
\begin{align}
\begin{split}
G_{\ell m}^{\rm CPM} &\sim r_p^{-2} + \mathcal{O} \left( r_p^{-3} \right), 
\ \
F_{\ell m}^{\rm CPM} \sim r_p^{-1} + \mathcal{O} \left( r_p^{-2} \right).
\end{split}
\end{align}

In the next section, by examining the ways in which the different variables are
related, I develop a general method for constructing master functions
with progressively higher-order large $r_p$ convergence.

\section{Time derivatives of master functions}
\label{sec:masterDeriv}

Beginning in this section I drop $\ell m$ indices for brevity,
although I do indicate FD quantities with a subscript $\omega$. 
I use the harmonic decomposition and MP notation introduced by 
Martel and Poisson \cite{MartPois05} with source term notation
from Hopper and Evans \cite{HoppEvan10}. Also, I
define the symbols $\la \equiv (\ell +2)(\ell - 1)/2$,
$\La \equiv \la + 3M / r$.

\subsection{Relationships between the known master functions}

\subsubsection{Odd parity}

The CPM master function is defined through the following
linear combination of odd-parity RW gauge 
MP amplitudes and their first derivatives,
\be
\Psi_{\rm CPM}^{(0)} (t,r) \equiv \frac{r}{\la} 
\left( \pa_r h_t  
- \pa_t  h_r - \frac{2}{r} h_{t} \right) .
\label{eq:masterOdd}
\ee
The superscript $(0)$ indicates the number of time derivatives of the 
CPM variable.
This is a $C^{-1}$ function (that is, 
it has a jump in its value at the particle's location, 
but is otherwise smooth), and so taking its time derivative
(indicated with a dot) yields
a distribution with a time-dependent Dirac delta at the particle's location,
\begin{align}
\begin{split}	
\dot \Psi_{\rm CPM}^{(0)} (t,r)
&=
\frac{r}{\la} 
\left( \pa_t \pa_r h_t  
- \pa_t^2  h_r - \frac{2}{r} \pa_t h_{t} \right) \\
&= \frac{2 f}{r} h_r - \frac{r_p f_p}{\la} p_r (t) \d \left( r - r_p \right),
\end{split}
\end{align}
where the second equality follows from the field equations (see \cite{HoppEvan10}).
Since we know the magnitude of the delta function exactly, we subtract it off,
defining a new master function which is also $C^{-1}$,
\be
\label{eq:CPMRW}
\Psi_{\rm CPM}^{(1)} (t,r) \equiv
\dot \Psi_{\rm CPM}^{(0)} + \frac{r_p f_p}{\la} p_r (t) \d \left( r - r_p \right)
= \frac{2 f}{r} h_r .
\ee
The superscript $(1)$ means ``the first time derivative of the CPM variable
with the singular part subtracted.''
This is exactly twice the original RW variable. Except for exactly at 
the location of the particle, it is precisely the time derivative of the CPM
variable. Therefore, the normalization coefficients for $\Psi^{(0)}_{\rm CPM}$
are related to those of $\Psi^{(1)}_{\rm CPM}$ by
\begin{align}
\label{eq:CCPM}
C^{(0),\pm}_{{\rm CPM}, \omega}
=
\frac{C^{(1),\pm}_{{\rm CPM}, \omega}}{-i \omega}.
\end{align}
which is valid for all $\o \ne 0$.
The $\o \to 0$ limit is subtle for unbound motion and is discussed at length
in Ref.~\cite{HoppCard17}.

\subsubsection{Even parity}
The ZM function is defined as
\begin{align}
\Psi_{\rm ZM}^{(0)} (t,r) 
\equiv \frac{r}{\la+1} \left[ K 
+ \frac{1}{ \La} 
\l f^2 h_{rr} - r f \pa_r K \r \right].
\end{align}
It, too, is $C^{-1}$, so taking its time derivative introduces a Dirac delta,
\begin{align}
&
\dot \Psi_{\rm ZM}^{(0)} (t,r) 
= \frac{r}{\la+1} \left[ \pa_t K 
+ \frac{1}{ \La} 
\l f^2 \pa_t h_{rr} - r f \pa_t \pa_r K \r \right] \notag
\\ 
\hspace{1ex} &
=
\frac{1}{\La} \l r \pa_t K - f h_{tr}  \r
+ \frac{r_p^2 f_p}{\La_p (\la + 1)} q_{tr} (t) \d \l r - r_p \r,
\end{align}
where again I have used the field equations.
After subtracting the singular term, define a new master function which is also
$C^{-1}$,
\begin{align}
\begin{split}
\Psi_{\rm ZM}^{(1)} (t,r) 
&\equiv
\dot \Psi_{\rm ZM}^{(0)} - \frac{r_p^2 f_p}{\La_p (\la + 1)} q_{tr} (t) 
\d \l r - r_p \r \\
&=
\frac{1}{\La} \l r \pa_t K - f h_{tr}  \r .
\end{split}
\end{align}
This is exactly the original Zerilli variable 
(although he wrote it in the FD). Except at the exact
location of the particle, it is precisely the time derivative of the ZM
variable. As before, note that the $\Psi_{\rm ZM}^{(0)}$ and $\Psi_{\rm ZM}^{(1)}$
normalization coefficients are related via
\begin{align}
\label{eq:CZM}
C^{(0),\pm}_{{\rm ZM}, \omega}
=
\frac{C^{(1),\pm}_{{\rm ZM}, \omega}}{-i \omega}.
\end{align}

The conclusion to draw from these variables is, \emph{taking the time 
derivative of a master functions and then removing the offending Dirac delta 
will always produce a new master function of the same form}.

\subsection{Master functions with an arbitrary number of time derivatives}

In the previous subsection, I showed that one can derive one master function from 
another by taking the time derivative and subtracting the exact delta function
that follows from differentiating a $C^{-1}$ function. Now I generalize that 
process to show how one can differentiate arbitrarily many times, each time 
creating a new master function of the same form.

Consider a master function $\Psi^{(0)}$ which satisfies an equation of the form
\begin{align}
\begin{split}
\label{eq:master1}
& \left[-\frac{\pa^2}{\pa t^2} + \frac{\pa^2}{\pa r_*^2} - V(r)\right]
\Psi^{(0)}(t,r) \\
&
\hspace{10ex}
= G^{(0)}(t) \d \l r - r_p \r 
+
F^{(0)}(t) \d ' \l r - r_p \r.
\end{split}
\end{align}
It is convenient to write the master function in the weak form
\be
\label{eq:weakPsi}
\Psi^{(0)}(t,r) = \Psi^{(0),+}(t,r) \th  \l r - r_p \r 
+
\Psi^{(0),-}(t,r) \th \l r_p - r \r .
\ee
The Heaviside coefficients satisfy the homogeneous version of Eq.~\eqref{eq:master1},
while the source terms imply 
that the jump in $\Psi^{(0)}$ at the particle's location is
\cite{HoppEvan10}
\begin{align}
\label{eq:psiJump}
\llbracket \Psi^{(0)} \rrbracket_{p} (t)
&= \frac{\mathcal{E}^2}{f_{p}^2 U_{p}^2} F^{(0)}.
\end{align}
$\mathcal{E}$ is the particle's specific energy and
$U_p^2$ is the effective potential, defined in App.~\ref{sec:rwzFull}.
Now, we can take the time derivative of 
any master function that satisfies Eq.~\eqref{eq:master1}
and it will satisfy an equation of the same form. However, the 
time derivative will introduce a second derivative of the delta function
in the source on the RHS. This follows because the time derivative 
of $\Psi^{(0)}$ is
\begin{align}
\begin{split}
\dot \Psi^{(0)}(t,r) =& 
\dot \Psi^{(0),+} \th  \l r - r_p \r 
+
\dot \Psi^{(0),-} \th \l r_p - r \r \\
&
-
\dot r_p \llbracket \Psi^{(0)} \rrbracket_{p}
\d \l r - r_p \r.
\end{split}
\end{align}
Thus, I define 
\begin{align}
\label{eq:psi1Def}
\Psi^{(1)} (t,r) &\equiv \dot \Psi^{(0)} + 
\frac{\dot r_p \mathcal{E}^2}{f_{p}^2 U_{p}^2} F^{(0)} 
\d \l r - r_p \r ,
\end{align}
which is $C^{-1}$
and will satisfy an equation of the form \eqref{eq:master1} with no
$\delta ''(r-r_p)$ source term.

\begin{widetext}

\subsection{Higher-order source terms}

It is fine to define new master functions by differentiating the old functions,
but what makes a master function unique is its source term. In order to 
find the source term for $\Psi^{(1)}$, start by acting with the wave
operator on Eq.~\eqref{eq:psi1Def},
\begin{align}
\label{eq:master2}
\left[-\frac{\pa^2}{\pa t^2} + \frac{\pa^2}{\pa r_*^2} - V(r)\right]
\left[ \dot \Psi^{(0)} + 
\frac{\dot r_p \mathcal{E}^2}{f_{p}^2 U_{p}^2} F^{(0)}
\d \left( r - r_p \right)\right]
= 
G^{(1)} \delta (r-r_p)
+
F^{(1)} \delta '(r-r_p) .
\end{align}
When the wave operator hits $\dot \Psi^{(0)}$, it is equivalent to
taking the time derivative of Eq.~\eqref{eq:master1} and so
\begin{align}
\begin{split}
\label{eq:master3}
&
\left[-\frac{\pa^2}{\pa t^2} + \frac{\pa^2}{\pa r_*^2} - V(r)\right]
\left[ 
\frac{\dot r_p \mathcal{E}^2}{f_{p}^2 U_{p}^2} F^{(0)}
\d \left( r - r_p \right)\right] 
 +
\dot G^{(0)} \delta (r-r_p) 
+
\left(\dot F^{(0)}-G^{(0)} \dot r_p\right) \delta '(r-r_p)\\
& \hspace{55ex}
-
F^{(0)} \dot r_p \delta ''(r-r_p) 
= 
G^{(1)} \delta (r-r_p)
+
F^{(1)} \delta '(r-r_p) .
\end{split}
\end{align}
Acting on the remaining delta term requires care.
The $r_*$ derivatives must be expanded as $\pa^2_{r_*} = f(r) \pa_r f(r) \pa_r$, and then
any functions of $r$ must be evaluated at $r = r_p(t)$ by using the identities
(for a smooth test function $g$)
\begin{align}
\begin{split}
\label{eq:deltaIdents}
g(x) \ \d (x) &= g (0) \ \d (x), \\
g(x) \ \d ' (x) &= g (0) \ \d ' (x)  - g'(0) \ \d (x), \\
g(x) \ \d '' (x) &= g (0) \ \d '' (x)  - 2 g'(0) \ \d ' (x) + g''(0) \ \d (x).
\end{split}
\end{align}
By design the $\d '' (r-r_p)$ term cancels. Then, equating the remaining 
coefficients of $\d (r-r_p)$ and $\d '(r-r_p)$ respectively yields
\begin{align}
\begin{split}
\label{eq:GExpr}
G^{(1)} =& \
\dot G^{(0)}
-\frac{\mathcal{E}^2 \dot r_p \ddot F^{(0)}}{f_p^2 U_p^2} 
+\frac{2 \dot F^{(0)}}{r_p^7 U_p^4} 
\bigg[
10 \mathcal{L}^4 M^2
- 
7 \mathcal{L}^4 M r_p
+
\left(\mathcal{L}^4+16 \mathcal{L}^2 M^2\right) r_p^2
+
2 \mathcal{L}^2 M \left(4 \mathcal{E}^2-5\right) r_p^3 \\
& 
+
\left(6 M^2+\mathcal{L}^2 -2 \mathcal{L}^2 \mathcal{E}^2\right) r_p^4 
+M \left(4 \mathcal{E}^2-3\right) r_p^5
\bigg] 
+
\frac{\dot r_p F^{(0)}}{r_p^{11} U_{p}^6}
\Bigg[
-
\frac{\mathcal{E}^2 r_p^{11} U_p^4 V_p}{f_{p}^2}
+
20 \mathcal{L}^6 M^3
-
30 \mathcal{L}^6 M^2 r_p \\ 
& \hspace{0ex}
+
4 \mathcal{L}^4 M\left(17 M^2+3 \mathcal{L}^2 \right) r_p^2 
-
\mathcal{L}^4 \left(\mathcal{L}^2+ 20 \mathcal{E}^2 M^2+102 M^2\right) r_p^3 
+
   \left(60 \mathcal{L}^2 M^3+ 8 \mathcal{L}^4 \mathcal{E}^2M +42 \mathcal{L}^4 M\right) 
r_p^4   \\ 
& \hspace{0ex}
-
2 \mathcal{L}^2 \left(45 M^2 -12 \mathcal{E}^2 M^2 
+ \mathcal{L}^2 \mathcal{E}^2+2\mathcal{L}^2\right) r_p^5 
+
12 M \left(M^2  -3 \mathcal{L}^2 \mathcal{E}^2+ 3 \mathcal{L}^2\right) r_p^6\\ 
& \hspace{0ex}
+
3 \left( 4 \mathcal{E}^2 M^2 - 6 M^2  +  2 \mathcal{L}^2 \mathcal{E}^2-\mathcal{L}^2\right) r_p^7 
+
6 M \left(1-2 \mathcal{E}^2\right) r_p^8 
\Bigg] 
\end{split} \\
\begin{split}
\label{eq:FExpr}
F^{(1)} =&
-
\dot r_p G^{(0)}
+ \frac{\dot F^{(0)}}{U_p^2 r_p^3} 
\Big[
2 \mathcal{L}^2 M
-
\mathcal{L}^2 r_p
+
2 M r_p^2
+
\l 2 \mathcal{E}^2-1 \r r_p^3
\Big]
+\frac{F^{(0)} \dot r_p}{r_p^7 U_p^4}
 \bigg[
-10 \mathcal{L}^4 M^2
+
7 \mathcal{L}^4 M r_p    \\
&
-
\left(\mathcal{L}^4+16 \mathcal{L}^2 M^2\right) r_p^2
+
10 \mathcal{L}^2 M \left(1-2 \mathcal{E}^2\right) r_p^3
-
\left(6 M^2+\mathcal{L}^2 -4 \mathcal{L}^2 \mathcal{E}^2\right) r_p^4 
+
3 M \left(1-4 \mathcal{E}^2\right) r_p^5\bigg]
\end{split}	
\end{align}
In writing these, I have used the constraints in 
Eq.~\eqref{eq:removeRDot} to reduce the number (and order of) time derivatives 
of $r_p$. Note the $V_p$ term in Eq.~\eqref{eq:GExpr}, which is the 
(parity-appropriate) potential evaluated at $r=r_p$.

The expressions \eqref{eq:GExpr} and \eqref{eq:FExpr} are generic relations
between the source terms of 
any two master functions of different `order'. It is a straightforward 
(if tedious) task to insert the $G$ and $F$ terms from the CPM (or ZM) variable on
the RHS and obtain the $G$ and $F$ terms from the RW  (or Z) variable.
It remains to show that the higher-order source terms converge more quickly
at large $r_p$ than the lower-order ones.

Expanding each of the terms on the RHS of Eq.~\eqref{eq:GExpr}
in inverse powers of $r_p$ confirms that $G^{(1)}$ is guaranteed
to fall off at least one power of $r_p$ faster than $G^{(0)}$.
Meanwhile, $F^{(1)}$ is guaranteed to converge at least as fast
as $F^{(0)}$. It is evident that if $F^{(1)}$ does not `out-converge'
$F^{(0)}$, then proceeding to the next order will produce a new source
$F^{(2)}$, which will converge faster. This exact phenomenon occurs in the 
right panel of Fig.~\ref{fig:GFConv} when  $F^{(3)}_{\rm ZM}$ converges
at the same rate as $F^{(2)}_{\rm ZM}$, but then $F^{(4)}_{\rm ZM}$
converges more rapidly.

Since the master functions $\Psi^{(1)}$ is essentially the time-derivative
of $\Psi^{(0)}$, the two functions have associated normalization 
coefficients related by
$C_{\omega}^{(0),\pm} = C_{\omega}^{(1),\pm}/(- i \omega).$
Each higher order brings one extra factor of $(- i \omega)^{-1}$. By
adjusting the normalization coefficients this way, one can always get 
back to the CPM and ZM variables which allow TD MP reconstruction.

\begin{figure*}[t]
\includegraphics[width=\textwidth]{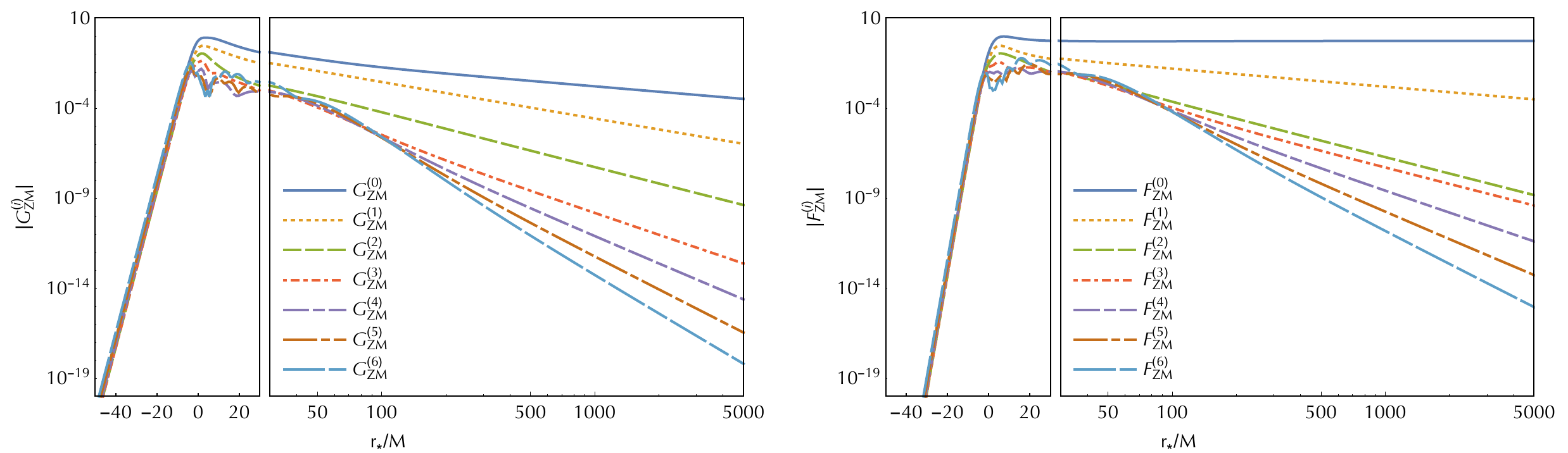}
\caption{
\label{fig:GFConv}
The $(\ell,m) = (2,2)$ mode source terms for a plunging particle with
$\mathcal{E} = 3$ and $\mathcal{L}_{\rm frac} = 0.99$ (see App.~\ref{sec:geos}
for orbit parametrization). 
In each of these graphs, the left side shows a log-linear plot,
emphasizing the exponential decay of the source terms at the horizon.
On the right of each graph, the horizontal axis changes to logarithmic, displaying
the algebraic convergence at large distance. Note that $F_{\rm ZM}^{(0)}$,
the ZM source, does not converge [see Eq.~\eqref{eq:ZMTrend}].
The next-order source term ($F_{\rm ZM}^{(1)}$) converges only
as $r_p^{-1}$, while the second-order term ($F_{\rm ZM}^{(2)}$)
jumps to an efficient $r_p^{-3}$ convergence. 
}
\end{figure*}

\end{widetext}

\begin{figure*}[t]
\includegraphics[width=\textwidth]{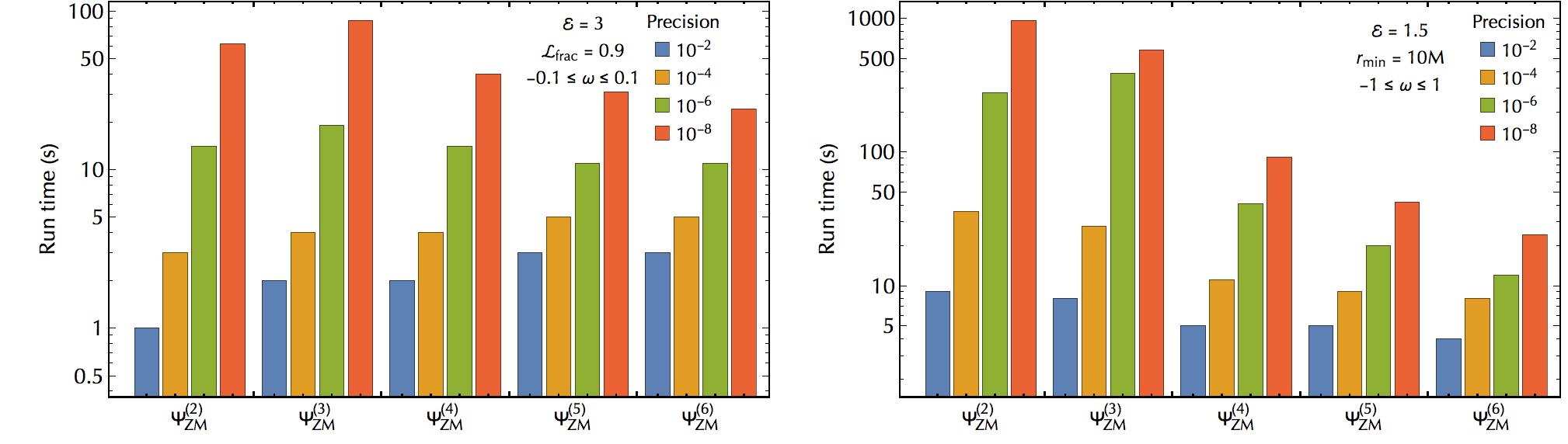}
\caption{
\label{fig:speedCharts}
Run times for a variety of precision requirements and 
master function choices. All runs are for the $(\ell, m) = (2,2)$ mode
and 20 evenly spaced harmonics in the shown $\omega$ range 
(excluding $\omega = 0$). The plots show that the benefit of
high-order master functions are realized when $\omega$ gets large.
See also discussion in the text.}
\end{figure*}

\section{Implementation and Results}
\label{sec:results}

\subsection{Numerical implementation}

I now briefly describe the algorithmic details of my numerical implementation. 
The code is written in $C$, uses a GSL integrator \cite{GSL},
and works in the following series of steps.

\begin{enumerate}
\item \emph{Solve geodesic equations.} A discussion of unbound geodesics
is given in App.~\ref{sec:rwzFull}.
\item \emph{Find boundary conditions to homogeneous FD master equation
\eqref{eq:FDmastereqn} for a given $\ell,m,\omega$ mode}. 
The infinity-side (out-going wave) solution is 
found with an asymptotic expansion in $(\omega r)^{-1}$ 
while the horizon-side (down-going wave) solution 
is found from a Taylor expansion in $f$.
This is equivalent to the way homogeneous
solutions are found in the bound case, e.g.~\cite{HoppEvan10}.
\item \emph{Integrate homogeneous solutions to source boundary}.
Numerically integrate both homogeneous solutions to the point
where the source is closest to the black hole. For scattering events 
this is the periapsis $r_{\rm min}$. For plunging trajectories,
usually around $r_* = -40M$ is plenty close (note the magnitude
of the source terms there in Fig.~\ref{fig:GFConv}).
\item \emph{Concurrently integrate homogeneous solutions with normalization
integral, \eqref{eq:EHSC}.}
Starting close to the black hole, numerically integrate out to a large $r_p$.
Then, double $r_p$, integrate again and check for convergence of 
Eq.~\eqref{eq:EHSC}. App.~\ref{sec:rwzFull} has a brief discussion of the 
best independent variables to use for this integral.
Note that for plunges, the horizon flux, and hence the $C^{-}_{\ell m \omega}$
integral is known to diverge, \cite{DaviRuffTiom72}. For the scattering
case, there are two (asymmetric, due to $\dot r_p$ terms) 
legs of the trajectory, which both must be covered in the source integration.
\item \emph{Repeat steps $2-4$ for a range of $\ell, m, \omega$}. 
While $m$ always ranges from $-\ell$ to $\ell$, both $\ell$ and $\omega$
have infinite range (with each $\omega$ spectrum being dense). The choice
of how to truncate (and discretize) 
these ranges rests on overall accuracy requirements.
\end{enumerate}

For fluxes, the 
$\ell$-sum consistently converges exponentially, so choosing
where to stop that infinite sum is straightforward. For a given $\ell,m$ though,
the relevant range of $\omega$ to choose is far from obvious. 
Weaker-field events radiate less and require a finer discretization of
the $\omega$ range. Further, the energy 
spectrum of any given $\ell,m$ mode exhibits
numerous `zeros' where the spectrum vanishes for a given $\omega$, only to
rise again beyond that point, making detecting spectral convergence
quite challenging. Thus, a fair amount of logic must be programmed into
any algorithm in order for it to dynamically determine how to truncate
these infinite sums.
The features of the energy spectrum discussed here can be seen in
the figures of the accompanying work Ref.~\cite{HoppCard17}.

A last challenge with the $\omega$-spectrum is the static mode
$\omega = 0$. Smarr \cite{Smar77} pointed out that the energy
spectrum goes to a constant in this limit whenever the particle's
speed is non-zero at infinity. 
It is simplest to consider 
very small modes, but skip the static mode itself.
We explore the zero-frequency limit thoroughly in \cite{HoppCard17}.

\subsection{Practical effects of higher-order master functions}

At first glance, there appears no downside to using higher-order master 
functions. Certainly, ever higher-order functions do provide ever-faster
converging source terms. However, in attempting to implement such a scheme,
 practical issues quickly arise. First, after the first two orders, 
the new source terms quickly become large, so large in fact that the
straightforward evaluation of them starts to outweigh the benefits of faster
$r_p$-convergence. To some extent, this trouble can be sidestepped by an
efficiently written code. It is best to precompute
position independent terms that show up in $G$ and $F$ 
before performing the normalization integral. Even then, the choice of 
which master function to use is not obvious, as 
can be seen in Fig.~\ref{fig:speedCharts}.

The figure shows two separate runs, performed for a range of precision 
requirements with various master functions.
Looking at the left panel, we see that when $\omega$ is relatively
small (in magnitude), there is little effect on run times from choosing a different
master function. For high precision requirements, the higher-order function
does lead to faster runs, but the benefit is minor. For low precision,
increasing the order of the master function actually \emph{worsens}
the runs times. The benefit of a higher-order master function becomes 
evident at larger frequencies. In the right panel of Fig.~\ref{fig:speedCharts},
the high-order master functions out-perform the lower-order functions
significantly, especially when precision requirements are high.
The most significant feature of this figure is what is not shown,
namely Zerilli's original variable $\Psi_{\rm ZM}^{(1)}$. 
As discussed, the source term for $\Psi_{\rm ZM}^{(1)}$ decays like $r_p^{-1}$,
which is slow enough to make numerical convergence to high precision impossible.

If one is not interested in results with precision beyond (say) $10^{-4}$
and only considering moderately large-$\omega$ (which excludes ultrarelativistic
events), the high-order master functions are usually unnecessary.
The new even-parity master function $\Psi_{\rm ZM}^{(2)}$ and the 
original odd-parity RW function $\Psi_{\rm CPM}^{(1)}$ provide a convenient 
`sweet-spot' of fast $r_p$-convergence with relatively compact source terms.
They are each given in App.~\ref{app:Sources}. For high-energy runs though,
the relevant range of harmonics gets high, and the benefit of higher-order
master functions is substantial.

\begin{figure*}
\includegraphics[width=\textwidth]{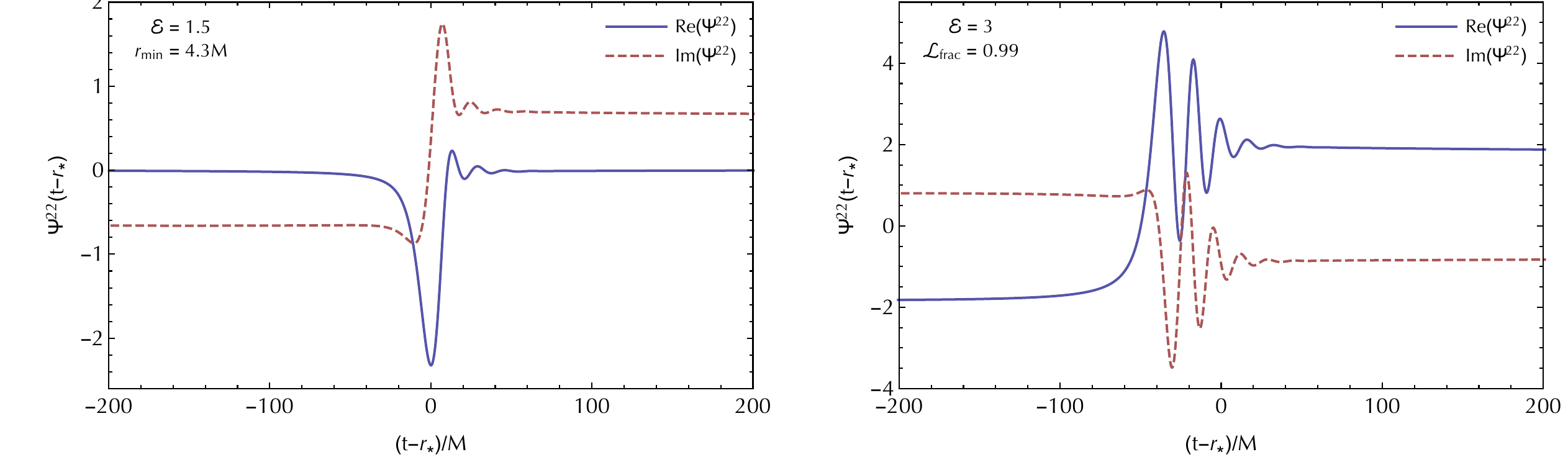}
\caption{
\label{fig:waveforms}
Example results from the code presented here. 
Both panels show the $(\ell,m) = (2,2)$ modes of 
the TD waveforms.
The left panel gives a sample waveform for a scattering run,
and the right considers a sample plunge. Many more results
from scattering runs computed with this code can be found in Ref.~\cite{HoppCard17}.
}
\end{figure*}

\section{Prospects for application to other formalisms and problems}
\label{sec:prospects}

The methods developed here to improve source behavior at large distances
are ideally suited to the RWZ formalism. It is worth exploring to what extent
the same techniques could be applied to other systems.

\subsection{Lorenz gauge on Schwarzschild}

The most natural extension of this technique would be to the Lorenz gauge 
field equations on Schwarzschild spacetime. As mentioned, there should
be no trouble using the equations in their natural form since each
of the source terms falls off as $1/r_p^2$ at large $r_p$. However, especially
for large Lorentz factors it is useful to have source terms which fall off
even faster. Without working out the details, I will show why we can 
probably increase the rate of large-$r$ source convergence.

In the TD, each of the (unconstrained) Lorenz gauge field equations is of the form
\begin{align}
\begin{split}
\label{eq:LorenzEqns31}
\l -\frac{\pa^2}{\pa t^2} + \frac{\pa^2}{\pa r_*^2} \r h^{(0)}_i (t,r) 
+ \mathcal{M}_{ij}  h^{(0)}_j (t,r) \\
= S^{(0)}_i (t) \d \l r - r_p \r,
\end{split}
\end{align}
where $h^{(0)}_i$ represents any of the 10 MP amplitudes 
($h_{tt}, h_{tr}, h_{rr}, j_t, j_r K, G, h_t, h_r, h_2$), 
$\mathcal{M}_{ij}$ is a matrix coupling the fields and
their first derivatives, and $S_i^{(0)} (t)$ are
the source amplitudes coming from the particle's stress energy projection.
The structure of the Lorenz gauge equations, with a wave operator on the
LHS and only a delta function (no delta prime) on the RHS implies that 
each of the amplitudes $h^{(0)}_i$ is $C^0$. 
Taking one time derivative yields
\begin{align}
\begin{split}
\label{eq:LorenzEqns21}
&\l -\frac{\pa^2}{\pa t^2} + \frac{\pa^2}{\pa r_*^2} \r \dot h^{(0)}_i (t,r) 
+ \mathcal{M}_{ij} \dot h^{(0)}_j (t,r) \\
&\hspace{5ex} = \dot S_i^{(0)} (t) \d \l r - r_p \r 
-
\dot r_p S_i^{(0)} (t) \d ' \l r - r_p \r,
\end{split}
\end{align}
The delta prime on the RHS implies that the fields $\dot h^{(0)}_i$ are 
$C^{-1}$ with jumps equal to
\begin{align}
\llbracket \dot h^{(0)}_i \rrbracket_{p} (t)= 
- \frac{\dot r_p \mathcal{E}^2}{f_{p}^2 U_{p}^2}  S_i^{(0)}.
\end{align}
Taking the second time derivative of the field equations gives
\begin{align}
\begin{split}
& \l -\frac{\pa^2}{\pa t^2} + \frac{\pa^2}{\pa r_*^2} \r \ddot h^{(0)}_i (t,r) 
+ \mathcal{M}_{ij}  \ddot h^{(0)}_j  (t,r) \\
&\hspace{5ex} = 
\ddot S_i^{(0)} \d \l r - r_p \r
- 
2 \dot r_p \dot S_i^{(0)} \d ' \l r - r_p \r \\
&\hspace{8ex}-
\ddot r_p S_i^{(0)}  \d ' \l r - r_p \r
+
\dot{r}^2_p S_i^{(0)}  \d ''\l r - r_p \r.
\end{split}
\end{align}
The $\d ''(r - r_p)$ source implies a $\d (r-r_p)$ term in each $\ddot h^{(0)}_i$ 
with a coefficient of $-\dot r_p \llbracket \dot h^{(0)}_i \rrbracket_{p}$.
Subtracting this term, we can define a new set of amplitudes 
\begin{align}
\label{eq:h2Def}
h^{(2)}_i (t,r) \equiv 
\ddot h^{(0)}_i (t,r)  
- \frac{\dot r_p^2 \mathcal{E}^2}{f_{p}^2 U_{p}^2}  S_i^{(0)} \d \l r - r_p \r,
\end{align}
which are all $C^{-1}$. They are exactly the second time derivative 
of the Lorenz gauge fields, except at the exact location of the particle.
These fields satisfy equations of the form
\begin{align}
\begin{split}
\label{eq:h2Eq}
&\l -\frac{\pa^2}{\pa t^2} + \frac{\pa^2}{\pa r_*^2} \r h^{(2)}_i  (t,r)
+ \mathcal{M}_{ij}  h^{(2)}_j  (t,r) \\
&
\hspace{5ex}
= 
S_i^{(2)} (t) \d \l r - r_p \r
+ 
R_i^{(2)} (t) \d ' \l r - r_p \r .
\end{split}
\end{align}
The differential operator on the LHS is precisely the Lorenz gauge operator.
It remains to act with Eq.~\eqref{eq:h2Eq} on the new fields \eqref{eq:h2Def}
to derive expressions for the new source terms $S_i^{(2)}$ and $R_i^{(2)}$.
The output is too lengthy and tedious to include here, but I have done so
and seen that each term decays at least as $r_p^{-3}$, indicating that
source integrations would converge more quickly by using 
$S_i^{(2)}$ and $R_i^{(2)}$ source terms rather than $S_i^{(0)}$.
Note that if this method were employed, the normalization procedure
laid out in Refs.~\cite{AkcaWarbBara13} and \cite{OsbuETC14} would have to 
be modified slightly in order to take into account the delta prime
in the source.
Lastly, it is worth remembering that the primary reason to use the Lorenz
gauge equations is for local GSF calculations, but this paper has not addressed
local calculations.

\subsection{Unbound motion on Kerr}

The situation on Kerr is less promising.
Perturbations on a Kerr background are typically found by solving
the Teukolsky equation. It is a wave operator in all four spacetime
variables acting on either of the Weyl scalars $\psi_4$ or $\psi_0$.
Schematically it is of the form
\begin{align}
\label{eq:TeukFull}
&\mathcal{W} \big[ \psi (t,r, \th, \vp) \big] = 
G(t) \d^3 \l x^i -x^i_p \r \\
& \hspace{2ex} + F_j(t) \pa_j \d^3 \l x^i -x^i_p \r
+ E_{jk}(t) \pa_j \pa_k \d^3 \l x^i -x^i_p \r, 
\notag
\end{align}
where the indices $i, j, k$ range over $r, \th, \vp$. 
One can exploit the axial symmetry of 
the Kerr background and reduce the $3+1$ Teukolsky equation
to a set (over azimuthal number $m$) 
of $2+1$ equations, at which point the schematic form is
\begin{align}
\label{eq:Teuk21}
&\mathcal{W}_m \big[ \psi_m (t,r, \th) \big] = 
 G^m(t) \d^2 \l x^a -x^a_p \r \\
& \hspace{2ex} 
+ F^m_b(t) \pa_b \d^2 \l x^a -x^a_p \r
+ E^m_{bc}(t) \pa_b \pa_c \d^2 \l x^a -x^a_p \r, 
\notag
\end{align}
where the indices $a, b, c$ range over $r, \th$. 

Unfortunately, the Teukolsky equation has never been
decomposed into a decoupled set of $1+1$ equations in $t$ and $r$.
In order to decompose it further, one must go all the way to
the FD and obtain an ODE in $r$ 
(as well as a separate homogeneous ODE in $\th$).
The method I have presented in this paper relies on
a $1+1$ wave equation and detailed knowledge of the weak structure
of the field at the particle's location. One can imagine
taking a similar approach to the $2+1$ equation \eqref{eq:Teuk21}, 
but problems quickly arise. The issue is that the distributions
on the RHS of Eq.~\eqref{eq:Teuk21} do not stem from
taking derivatives of Heavisides and Dirac deltas. Rather, they
are due to the local $1/r$ divergence of the particle's own field.
An attempt to write down a $2+1$ weak form of $\psi_m (t, r, \th)$ akin
to that in Eq.~\eqref{eq:weakPsi} succeeds only in the sense that
the coefficients of the Heavisides are not finite at $(r,\th) = (r_p, \th_p)$.
Indeed, this very divergence is at the heart of all GSF research
over the last two decades.

It seems then that the best bet for applying the methods developed 
in this paper would be to decompose Eq.~\eqref{eq:Teuk21} into a
\emph{coupled} set of 1+1 equations. Clearly, this is not ideal,
and it is worth admitting that the Sasaki-Nakamura formalism is probably
a far simpler and more effective way to study unbound motion on Kerr.

\acknowledgments
I thank Vitor Cardoso for helpful discussions and Thomas Osburn
for use of his numerical integrator.
I acknowledge financial support provided under the European Union's H2020 ERC Consolidator Grant ``Matter and strong-field gravity: New frontiers in Einstein's theory'' grant agreement no. MaGRaTh--646597.
This article is based upon work from COST Action CA16104 ``GWverse'', supported by COST (European Cooperation in Science and Technology).
I thankfully acknowledge the computer resources, technical expertise and assistance provided by CENTRA/IST. Computations were performed at the cluster
``Baltasar-Sete-S\'ois,'' and supported by the MaGRaTh--646597 ERC Consolidator Grant.

\appendix

\section{The RWZ formalism}
\label{sec:rwzFull}

\subsection{Unbound Schwarzschild geodesics}
\label{sec:geos}
Consider a point particle of mass $\mu$ moving on a Schwarzschild background
of mass $M$.
Let the worldline be parametrized by proper time $\tau$, 
i.e.~$x_p^{\mu}(\tau) = \left[ t_p(\tau),r_p(\tau),\pi/2,\vp_p(\tau) \right]$,
where I have confined the particle to $\th_p = \pi/2$ without loss of generality. 
Generic geodesics are parametrized by  the specific energy $\mathcal{E}$ 
and the specific angular momentum $\mathcal{L}$. 
The four-velocity  $u^{\mu} = dx_p^{\mu}/d\tau$ can be written in terms of them as
\be
\label{eq:fourVelocity}
u^t = \frac{\mathcal{E}}{f_{p}},  
\q \q
u^{\varphi} = \frac{{\cal{L}}}{r_p^2} ,
\q \q
\l u^r \r^2 = \mathcal{E}^2-U^2_{p}, 
\ee
where the effective potential is
\be
\label{eq:effV}
U^2(r,\mathcal{L}^2) \equiv f \l 1 + \frac{\mathcal{L}^2}{r^2} \r .
\ee
The radial coordinate velocity obeys the constraints
\begin{align}
\begin{split}	
\label{eq:removeRDot}
\dot r_p^2(t) &= f_{p}^2 - \frac{f_{p}^2}{{\cal{E}}^2} U^2_{p} , 
\\
\ddot r_p (t) &=  \frac{2M f_{p}}{r_{p}^2} - \frac{f_{p}^2}{{\cal{E}}^2 r_{p}^2} 
\left[3 M -  \frac{{\cal{L}}^2}{r_{p}} 
+ \frac{5 M {\cal{L}}^2}{r_{p}^2} \right].
\end{split}
\end{align}

\subsubsection{Plunges}
Geodesics representing particles plunging from infinity 
must have $\mathcal{E} \ge 1$ and also clear the peak of the effective
potential, 
\begin{align}
U^2_{\rm max} 
=
\frac{1}{54} \left[\frac{\mathcal{L}^2}{M^2}
+36
+\sqrt{1-\frac{12 M^2}{\mathcal{L}^2}} \left(\frac{\mathcal{L}^2}{M^2}-12\right)\right].
\end{align}
Define the maximum value of specific angular momentum $\mathcal{L}_{\rm max}$
for a given $\mathcal{E}$ by solving $\mathcal{E}^2 = U^2_{\rm max}$.
Then, it is convenient to parametrize
such trajectories using  $\mathcal{E}$ and $\mathcal{L}_{\rm frac}$
where $\mathcal{L}_{\rm frac} \equiv \mathcal{L} / \mathcal{L}_{\rm max}$
ranges between $-1$ and $1$, exclusive, with $\mathcal{L}_{\rm frac} = 0$
corresponding to a head-on plunge.
Given $\mathcal{E}$ and $\mathcal{L}_{\rm frac}$ 
parametrizing a plunging geodesic, 
it is convenient to integrate the geodesic equations
and the normalization integral, \eqref{eq:EHSC} with respect to the tortoise 
coordinate $r_*$ since it approaches the horizon asymptotically.

\subsubsection{Scatters}

For scattering geodesics, it is better to replace $\mathcal{L}$
with either the periapsis $r_{\rm min}$ or the impact parameter $b$.
The former is related to $\mathcal{E}$ and $\mathcal{L}$ by solving
$U^2(r_{\rm min},\mathcal{L}) = \mathcal{E}^2$ for $r_{\rm min}$,
while the latter is defined as
$ b \equiv \mathcal{L} / \sqrt{\mathcal{E}^2 - 1}.$
In the special case of $\mathcal{E} = 1$ (parabolic motion), 
$b \to \infty$, making $r_{\rm min}$ preferable.

In addition to these two parameters, it is worth noting the natural
extension of the semilatus rectum $p$ and eccentricity $e$ to unbound
geodesics. As in bound motion, they obey the relations
\be
{\cal{E}}^2 = \frac{(p-2-2e)(p-2+2e)}{p(p-3-e^2)}, 
\ \
{\cal{L}}^2 = \frac{p^2 M^2}{p-3-e^2},
\ee
where now $e \ge 1$.
The $p,e$ parametrization, can be used 
with Darwin's \cite{Darw59} relativistic anomaly $\chi$,
and the radial position is 
\be
r_p \l \chi \r = \frac{pM}{1+ e \cos \chi} .
\ee
For eccentric motion $\chi$ runs from $0 \to 2\pi$ during 
one radial libration, but here $\chi$ ranges between 
$-\chi_{\infty} \to \chi_{\infty}$
where $\chi_{\infty} \equiv \arccos (-1/e)$, with
periapsis occurring at $\chi = 0$.
The particle's coordinate time $t_p$ is related to $\chi$ by
the first-order differential equation,
\begin{align}
\begin{split}
\label{eqn:darwinEqns}
\frac{dt_p}{d \chi} &= 
 \frac{p^2 M}{(p - 2 - 2 e \cos \chi) (1 + e \cos \chi)^2} \\
& \hspace{15ex}
\times
\left[ \frac{(p-2)^2 - 4 e^2}{p - 6 - 2 e \cos \chi} \right]^{1/2},
\end{split}
\end{align}
while $\vp_p$ is known analytically,
\be
\vp_p(\chi) = \left(\frac{4 p}{p - 6 - 2 e}\right)^{1/2} \, 
F\left(\frac{\chi}{2} \, \middle| \, -\frac{4 e}{p - 6 - 2 e}  \right) .
\ee
$F(x|m)$ is the incomplete elliptic integral of the first kind 
\cite{GradETC07}. 

When performing the normalization integral \eqref{eq:EHSC},
$\chi$ is a good parameter to use near periapsis, as it
removes troublesome $1/\dot r_p$ terms. However, further from
the encounter it is advantageous to switch to another curve parameter
like $t$ or $r_p$ to avoid having to take ever-smaller steps in $\chi$.

\subsection{Frequency domain formalism}

In Sec.~\ref{sec:rwzBackground} I briefly covered the RWZ formalism 
in the FD. The TD master equation \eqref{eq:TDmastereqn} and its FD 
counterpart are connected by the spectral decomposition of the field
$\Psi_{\ell m}$ and the source 
$S_{\ell m}$,
\begin{align}
\begin{split}
\label{eq:fourierTransf}
\Psi_{\ell m}(t,r) &= \frac{1}{2\pi} \int_{-\infty}^\infty 
X_{\ell m\o}(r) \, e^{-i \o t}  d\o , 
\\
S_{\ell m}(t,r) &= \frac{1}{2\pi} \int_{-\infty}^\infty 
Z_{\ell m\o}(r) \, e^{-i \o t}  d\o .
\end{split}
\end{align}
Formally, the Fourier coefficients are found by integrating over all time,
\begin{align}
\begin{split}
\label{eq:invFourierTransf}
X_{\ell m\o}(r) &= \int_{-\infty}^\infty 
\Psi_{\ell m}(t,r) \, e^{i \o t}  dt , 
\\
Z_{\ell m\o}(r) &= \int_{-\infty}^\infty 
S_{\ell m}(t,r) \, e^{i \o t}  dt .
\end{split}
\end{align}

Retarded boundary conditions require the two desired independent homogeneous
solutions to Eq.~\eqref{eq:FDmastereqn} to behave as
\begin{align}
\begin{split}
\hat X_{\ell m\o}^+ (r_* \to +\infty)  &\sim e^{i \o r_*}, \\
\hat X_{\ell m\o}^- (r_* \to -\infty)  &\sim e^{-i \o r_*}.
\end{split}
\end{align}
A Green function is formed from these two solutions 
and integrated over the source function $Z_{\ell m\o}(r)$ to obtain the 
particular solution of Eq.~\eqref{eq:FDmastereqn},
\be
\label{eq:FDInhomog}
X_{\ell m\o} (r) = c^+_{\ell m\o}(r) \, \hat{X}^+_{\ell m\o}(r)
+ c^-_{\ell m\o}(r) \, \hat{X}^-_{\ell m\o}(r) ,
\ee
where the normalization functions are given by the integrals
\begin{align}
\label{eq:cPM}
\begin{split}
c^+_{\ell m\o}(r) &= \frac{1}{W_{\ell m\o}} \, \int_{2M}^r   
\frac{dr'}{f(r')} \hat{X}^-_{\ell m\o}(r') \, Z_{\ell m\o}(r') , 
\\
c^-_{\ell m\o}(r) &= \frac{1}{W_{\ell m\o}} \, \int_r^{\infty} 
\frac{dr'}{f(r')} \hat{X}^+_{\ell m\o}(r') \, Z_{\ell m\o}(r') .
\end{split}
\end{align}
Here $W_{\ell m\o}$ is the (constant in $r_*$) Wronskian 
\be
W_{\ell m\o} = f(r) \left( \hat{X}^-_{\ell m\o} \frac{d \hat{X}^+_{\ell m\o}}{dr}
- \hat{X}^+_{\ell m\o} \frac{d \hat{X}^-_{\ell m\o}}{dr} \right) .
\ee
The limits of integration in 
\eqref{eq:cPM} are extended to cover all values of $r$, yielding the 
normalization coefficients
\be
\label{eq:normC}
C_{\ell m\o}^{\pm} 
= \frac{1}{W_{\ell m\o}} \int_{2M}^{\infty} dr
 \ \frac{\hat X^{\mp}_{\ell m\o} (r) Z_{\ell m\o} (r)}{ f(r)} .
\ee
In practice this integral is solved by substituting $Z_{\ell m \omega}$
from Eq.~\eqref{eq:invFourierTransf} into Eq.~\eqref{eq:normC} from which follows
Eq.~\eqref{eq:EHSC}. As shown in Ref.~\cite{HoppCard17} these coefficients 
can be used
to compute radiated energy as well as harmonics of the gauge invariant waveform.

\begin{widetext}

\section{Explicit expressions for $\Psi_{\rm ZM}^{(1)}$,
$\Psi_{\rm ZM}^{(2)}$ and $\Psi_{\rm CPM}^{(1)}$
sources}
\label{app:Sources}
Explicit ZM and CPM source terms are given in Ref.~\cite{HoppEvan10}. Here I provide 
expressions for the next two even-parity source terms 
and the next one odd-parity source term.
The first time derivative of the ZM function is the Zerilli function. Its source term is
\begin{align}
\begin{split}
G^{(1)}_{\rm ZM}
&=
\bar Y_{\vp \vp} \left\{-\frac{8 \pi  f_p \mathcal{L}^2 \mu  \dot r_p \left[30 M^2+3 (4 \lambda -3) M r_p-4 \lambda  r_p^2\right]}{\lambda  (\lambda +1) \La_p
   r_p^6 \mathcal{E}}  
   +\frac{8 i \pi  f_p^3 \mathcal{L}^3 \mu  m}{\lambda  (\lambda +1) r_p^5 \mathcal{E}^2}\right\} \\
&
+\frac{8 \pi  f_p^2 \mathcal{L} \mu  \bar Y_{\vp}}{(\lambda +1) \La_p^2 r_p^8 \mathcal{E}^2}
   \Big\{ 54 \mathcal{L}^2 M^3
   +
   \mathcal{L}^2 (28 \lambda -33) M^2 r_p
   +
   M r_p^2 \left[\mathcal{L}^2 \left(2 \lambda ^2-18 \lambda +3\right)+42 M^2\right]
   \\
& + 
   r_p^3 \left[ 3 M^2 \left(8 \lambda +2 \mathcal{E}^2-7\right)-\mathcal{L}^2 (\lambda -2) \lambda \right]
      +
   2 \lambda  M r_p^4 \left(\lambda +3 \mathcal{E}^2-7\right)
   -
   \lambda  r_p^5 \left(\lambda +2 \mathcal{E}^2-1\right)
   \Big\}
   \\
&
   -\frac{8 \pi  f_p \mu  \dot r_p \bar Y}{\La_p^2 r_p^6 \mathcal{E}}
    \Big[ 18 \mathcal{L}^2 M^2
   +
   \mathcal{L}^2 (8 \lambda -3) M r_p
    +
    2 r_p^2 \left(6 M^2-\mathcal{L}^2 \lambda \right)
   +
   6 \lambda  M r_p^3
   +
   \lambda  r_p^4 \left(2 \mathcal{E}^2-1\right)\Big] ,
   \end{split}
\end{align}
\begin{align}
F^{(1)}_{\rm ZM}
&=
\frac{8 \pi  f_p^2 \mathcal{L}^2
   \mu  \dot r_p \bar Y_{\vp \vp}}{\lambda  (\lambda +1) r_p^3 \mathcal{E}}
+
\frac{8 \pi  f_p^4 \mathcal{L} \mu  \bar Y_{\vp} \left(\mathcal{L}^2+r_p^2\right)}{(\lambda +1) \La_p r_p^4 \mathcal{E}^2}
+
\frac{8 \pi  f_p^2 \mu  \dot r_p \bar Y
   \left(\mathcal{L}^2+r_p^2\right)}{\La_p r_p^3 \mathcal{E}}.
\end{align}
The source for the second time derivative of the ZM function is
\begin{align}
& G_{\rm ZM}^{(2)}
=
-\frac{8 \pi  f_p^2 \mu  \bar{Y}}{\Lambda_p ^2 r_p^8 \mathcal{E}}
 \Big[ 18 \mathcal{L}^2 M^3+3 r_p^3 \left(2 M^2 \left(2 \lambda
   +\mathcal{E}^2-1\right)-\mathcal{L}^2 \lambda ^2\right)
+
   6 M r_p^2 \left(\mathcal{L}^2 (\lambda -2) \lambda +M^2\right) 
\notag
\\
& + 
2 \mathcal{L}^2 (11 \lambda -6) M^2 r_p+\lambda  M r_p^4
   \left(4 \lambda +6 \mathcal{E}^2-7\right)+2 \lambda ^2 r_p^5 \left(\mathcal{E}^2-1\right)\Big]
-
   \frac{16 \pi  f_p \mathcal{L} \mu  M \dot{r}_p
   \bar{Y}_{\vp} (12 M+(3 \lambda -2) r_p)}{(\lambda +1) \Lambda_p  r_p^6} 
\notag
\\
& +
\bar{Y}_{\vp \vp} \Bigg\{
\frac{16 i \pi  f_p
   \mathcal{L}^3 \mu  m \dot{r}_p \left(7 \mathcal{L}^2 M+11 M r_p^2-4 r_p^3-2 \mathcal{L}^2 r_p\right)}{\lambda  (\lambda +1) r_p^5
   \left(\mathcal{L}^2+r_p^2\right)^2}
+
\frac{8 \pi  f_p^2 \mathcal{L}^2 \mu}{\lambda  (\lambda +1) \Lambda_p ^2 r_p^9 \mathcal{E} \left(\mathcal{L}^2+r_p^2\right)^3} 
\Big[ 
252 \mathcal{L}^6 M^4
\\
&
-9 \mathcal{L}^6 M^3 r_p \left(-20 \lambda +2 m^2+19\right)
+
3 \mathcal{L}^4 M^2 r_p^2 \left(\mathcal{L}^2 \left(10 \lambda ^2-40 \lambda +(3-4 \lambda ) m^2+9\right)+360 M^2\right)
\notag
\\
& +
   \mathcal{L}^4 M r_p^3 \left(9 M^2 \left(84 \lambda -4 m^2+8 \mathcal{E}^2-81\right)-2 \mathcal{L}^2 \lambda  \left(\lambda ^2+11 \lambda +(\lambda -3)
   m^2-9\right)\right)
\notag
\\
&+r_p^4 \left(\mathcal{L}^6 \lambda ^2 \left(\lambda +m^2+4\right)+6 \mathcal{L}^4 M^2 \left(3 \left(7 \lambda ^2-28 \lambda +6\right)+(3-4 \lambda ) m^2+(7 \lambda -3)
   \mathcal{E}^2\right)+1620 \mathcal{L}^2 M^4\right)
\notag
\\
& +
\mathcal{L}^2 M r_p^5 \left(9
   M^2 \left(124 \lambda -2 m^2+32 \mathcal{E}^2-129\right)-2 \mathcal{L}^2 \lambda  \left(3 \left(\lambda ^2+15 \lambda -12\right)+2 (\lambda -3) m^2-2 (\lambda -3)
   \mathcal{E}^2\right)\right)
\notag
\\
&+
r_p^6 \left(3 \mathcal{L}^2 M^2 \left(62
   \lambda ^2-264 \lambda +(3-4 \lambda ) m^2+6 (10 \lambda -3) \mathcal{E}^2+63\right)+\mathcal{L}^4 \lambda ^2 \left(3 (\lambda +5)+2 m^2-2 (\lambda +2) \mathcal{E}^2\right)+792
   M^4\right)
\notag
\\
& +
M r_p^7 \left(9 M^2 \left(60 \lambda +40
   \mathcal{E}^2-67\right)-2 \mathcal{L}^2 \lambda  \left(3 \left(\lambda ^2+23 \lambda -21\right)+(\lambda -3) m^2-6 (2 \lambda -3) \mathcal{E}^2\right)\right)
\notag
\\
&
+
r_p^8 \left(\mathcal{L}^2 \lambda ^2 \left(3 \lambda
   +m^2-2 (2 \lambda +5) \mathcal{E}^2+24\right)+6 M^2 \left(15 \lambda ^2-68 \lambda +3 (13 \lambda -6) \mathcal{E}^2+18\right)\right)
\notag
\\
&
-2 \lambda  M r_p^9 \left(\lambda ^2+35 \lambda -18 (\lambda -2) \mathcal{E}^2-36\right)+\lambda ^2 r_p^{10}
   \left(\lambda -2 (\lambda +7) \mathcal{E}^2+13\right)\Big] \Bigg\},
\notag
\end{align}

\begin{align}
\begin{split}
F_{\rm ZM}^{(2)}
=&
-\frac{8 \pi  f_p^3 \mu  M \bar{Y} \left(\mathcal{L}^2+r_p^2\right)}{\Lambda_p  r_p^5 \mathcal{E}}+\frac{16 \pi  f_p^2 \mathcal{L} \mu  \dot{r}_p \bar{Y}_{\vp}}{(\lambda
   +1) r_p^3}
+
\bar{Y}_{\vp \vp} 
\Bigg\{-\frac{8 \pi  f_p^3 \mathcal{L}^2 \mu}{\lambda  (\lambda +1) \Lambda_p  r_p^6 \mathcal{E} \left(\mathcal{L}^2+r_p^2\right)^2}
\Big[21 \mathcal{L}^4 M^2\\
&
+
9 \mathcal{L}^4 (\lambda -1) M r_p 
+
r_p^2 \left(60 \mathcal{L}^2 M^2-4 \mathcal{L}^4 \lambda \right)
+
3\mathcal{L}^2 M r_p^3 \left(8 \lambda +2 \mathcal{E}^2-9\right)
+
r_p^4 \left(39 M^2+\mathcal{L}^2 \lambda  \left(2
   \mathcal{E}^2-11\right)\right)
\\
& 
+
3 M r_p^5 \left(5 \lambda +6 \mathcal{E}^2-6\right)
+
\lambda  r_p^6 \left(6 \mathcal{E}^2-7\right)\Big] 
-
\frac{16 i \pi  f_p^2 \mathcal{L}^3 \mu  m \dot{r}_p}{\lambda  (\lambda +1) r_p^3
   \left(\mathcal{L}^2+r_p^2\right)}\Bigg\}.
\end{split}
\end{align}

The time derivative of the CPM variable is the RW variable
(up to a factor of 2).
The RW source is
\begin{align}
G^{(1)}_{\rm CPM}
=
-\frac{8 \pi  f_p^2 \mathcal{L}^2 \mu  \bar X_{\vp \vp}}{\lambda  (\lambda +1) r_p^4 \mathcal{E}}-\frac{16 \pi  f_p \mathcal{L} \mu  \dot r_p \bar X_{\vp}}{(\lambda +1)
   r_p^3},
   \q \quad
F^{(1)}_{\rm CPM}
=
\frac{8 \pi  f_p^3 \mathcal{L}^2 \mu  \bar X_{\vp \vp}}{\lambda  (\lambda +1) r_p^3 \mathcal{E}}.
\end{align}
It is concise enough and
 converges fast enough to make it the ideal choice for most calculations. 
The next-order source is much longer and converges only as fast
as the RW source, so I do not include it here. It is only by proceeding to
the subsequent order and finding the $\Psi_{\rm CPM}^{(3)}$ source that a
faster convergence can be achieved.

\end{widetext}

\bibliography{Unbound}

\end{document}